\documentclass[12pt]{article}
\usepackage{epsfig}

\newcommand\pubnumber{SLAC-PUB-8763 \\LBNL-47226}
\newcommand\pubdate{January, 2001}
\newcommand\hepnumber{hep-ph/0101342}

\def\wellsad{Department of Physics, University of California, 
             Davis, CA 95616, and\\
       Lawrence Berkeley National Laboratory, Berkeley, CA 94720}
\def\wellsack{\footnote{Work supported by the Department of Energy
and the Alfred P. Sloan Foundation.}}
\def\SLAC{Stanford Linear Accelerator Center\\
    Stanford University, Stanford, California 94309 USA}
\def\doeack{\footnote{Work supported by the Department of Energy,
                     contract DE--AC03--76SF00515.}}

\def\Title#1{\begin{center} {\Large #1 } \end{center}}
\def\Author#1{\begin{center}{ \sc #1} \end{center}}
\def\Address#1{\begin{center}{ \it #1} \end{center}}
\def\andauth{\begin{center}{and} \end{center}}
\def\submit#1{\begin{center}Submitted to {\sl #1} \end{center}}
\newcommand\pubblock{\rightline{\begin{tabular}{l} \pubnumber\\
         \pubdate \\ \hepnumber  \end{tabular}}}
\newenvironment{Abstract}{\begin{quotation} \begin{center}
                       ABSTRACT
     \end{center}\bigskip  }{\end{quotation}}

\def\submit#1{\begin{center}Submitted to {\sl #1} \end{center}}
\def\Acknowledgements{\bigskip  \bigskip \begin{center} \begin{large}
             \bf ACKNOWLEDGEMENTS \end{large}\end{center}}
\def\beq{\begin{equation}}
\def\eeq#1{\label{#1}\end{equation}}
\def\eeqn{\end{equation}}

\newenvironment{Eqnarray}%
   {\arraycolsep 0.14em\begin{eqnarray}}{\end{eqnarray}}
\def\beqa{\begin{Eqnarray}}
\def\eeqa#1{\label{#1}\end{Eqnarray}}
\def\eeqan{\end{Eqnarray}}
\def\CR{\nonumber \\ }

\def\leqn#1{(\ref{#1})}

\let\bar=\overbar

\def\etal{{\it et al.}}

\def\VEV#1{\left\langle{ #1} \right\rangle}

\def\lsim{\mathrel{\raise.3ex\hbox{$<$\kern-.75em\lower1ex\hbox{$\sim$}}}}
\def\gsim{\mathrel{\raise.3ex\hbox{$>$\kern-.75em\lower1ex\hbox{$\sim$}}}}

\def\L{{\cal L}}

\def\half{\frac{1}{2}}

\def\Dslash{\not{\hbox{\kern-4pt $D$}}}
\def\dslash{\not{\hbox{\kern-2pt $\del$}}}

\def\ee{e^+e^-}
\def\sstw{\sin^2\theta_w}

\def\mz{m_Z}

\def\mw{m_W}

\def\msb{{\bar{\ssstyle M \kern -1pt S}}}

\makeatletter
\def\section{\@startsection{section}{0}{\z@}{5.5ex plus .5ex minus
 1.5ex}{2.3ex plus .2ex}{\large\bf}}
\def\subsection{\@startsection{subsection}{1}{\z@}{3.5ex plus .5ex minus
 1.5ex}{1.3ex plus .2ex}{\normalsize\bf}}
\def\subsubsection{\@startsection{subsubsection}{2}{\z@}{-3.5ex plus
-1ex minus  -.2ex}{2.3ex plus .2ex}{\normalsize\sl}}

\renewcommand{\@makecaption}[2]{%
   \vskip 10pt
   \setbox\@tempboxa\hbox{\small #1: #2}
   \ifdim \wd\@tempboxa >\hsize     % IF longer than one line:
       \small #1: #2\par          %   THEN set as ordinary paragraph.
     \else                        %   ELSE  center.
       \hbox to\hsize{\hfil\box\@tempboxa\hfil}
   \fi}

 \def\citenum#1{{\def\@cite##1##2{##1}\cite{#1}}}
\newcount\@tempcntc
\def\@citex[#1]#2{\if@filesw\immediate\write\@auxout{\string\citation{#2}}\fi
  \@tempcnta\z@\@tempcntb\m@ne\def\@citea{}\@cite{\@for\@citeb:=#2\do
    {\@ifundefined
       {b@\@citeb}{\@citeo\@tempcntb\m@ne\@citea\def\@citea{,}{\bf ?}\@warning
       {Citation `\@citeb' on page \thepage \space undefined}}%
    {\setbox\z@\hbox{\global\@tempcntc0\csname b@\@citeb\endcsname\relax}%
     \ifnum\@tempcntc=\z@ \@citeo\@tempcntb\m@ne
       \@citea\def\@citea{,}\hbox{\csname b@\@citeb\endcsname}%
     \else
      \advance\@tempcntb\@ne
      \ifnum\@tempcntb=\@tempcntc
      \else\advance\@tempcntb\m@ne\@citeo
      \@tempcnta\@tempcntc\@tempcntb\@tempcntc\fi\fi}}\@citeo}{#1}}
\def\@citeo{\ifnum\@tempcnta>\@tempcntb\else\@citea\def\@citea{,}%
  \ifnum\@tempcnta=\@tempcntb\the\@tempcnta\else
  {\advance\@tempcnta\@ne\ifnum\@tempcnta=\@tempcntb \else\def\@citea{--}\fi
    \advance\@tempcnta\m@ne\the\@tempcnta\@citea\the\@tempcntb}\fi\fi}
\makeatother

\def\seff{{\sin^2\theta_w^{\rm eff}}}
\begin{document}
\begin{titlepage}
\pubblock

\vfill
\Title{How Can a Heavy Higgs Boson be Consistent with the 
         Precision Electroweak Measurements?}
\vfill
\Author{Michael E. Peskin\doeack}
\Address{\SLAC}
\smallskip
\andauth
\smallskip
\Author{James D. Wells\wellsack}
\Address{\wellsad}
\vfill
\begin{Abstract}
The fit of precision electroweak data to the Minimal Standard Model 
currently gives an
upper limit on the Higgs boson mass of 170 GeV at 95\% confidence.
Nevertheless, it is often said that the Higgs boson could be much heavier 
in more general models.  In this paper, we critically review models that
have been proposed in the literature that allow a heavy Higgs boson 
consistent with the precision electroweak constraints.  All have unusual
features, and all can be distinguished from the Minimal Standard Model either
by improved precision measurements or by other signatures accessible to 
next-generation colliders.
\end{Abstract}
\vfill
\submit{Physical Review D}
\vfill
\end{titlepage}
\def\thefootnote{\fnsymbol{footnote}}
\setcounter{footnote}{0}
\tableofcontents
\newpage
\section{Introduction}
Is there a Higgs boson?  What is its mass?  These are among the most pressing
questions of contemporary elementary particle physics.  The final months of 
experiments at the LEP
$\ee$ collider at CERN showed tantalizing hints of the appearance of
the Higgs boson.  But with the LEP run now ended, 
we will not see further experimental evidence to confirm or refute
these suggestions for many years.  Thus, it is important to revisit 
the indirect constraints on the Higgs boson mass and to understand
their power as well as possible.

The most important indirect information on the Higgs boson comes from
precision measurements of the weak interactions \cite{LEWWG}.
The Minimal Standard 
Model (MSM)---defined as the $SU(3)\times SU(2) \times U(1)$ gauge theory of 
quarks and 
leptons with a single elementary Higgs field to break the electroweak 
symmetry---provides a good fit to the corpus of precision electroweak 
data.
The fit presented at the most recent International Conference on High-Energy
Physics predicts the mass of the Higgs boson to be less than 170 GeV
at 95\% confidence~\cite{Gurtu}.   Though this limit may be weakened slightly
by improved measurements of the renormalization of $\alpha$ \cite{Strom},
it remains true that the MSM with a  Higgs boson of mass above 250 GeV 
is strongly inconsistent with the current data.

On the other hand, it is likely that the correct picture of electroweak 
symmetry breaking requires ingredients beyond the MSM. In principle, these
new ingredients could affect the electroweak fit and weaken the upper limit
on the Higgs boson mass.  Specific models have been presented in which there
is no significant upper limit.  In this paper, we will review and catalogue
models with new physics beyond the MSM which allow a heavy Higgs boson
to be consistent with the precision electroweak measurements.  We will give
strategies for producing such models, and we will investigate what properties
these models share.

The predictions of the MSM depend on the Higgs boson mass through loop 
diagrams which contain the Higgs boson as a virtual particle.  A general
model of electroweak symmetry breaking might not contain a Higgs boson
as a light, narrow resonance.  However, any such model must contain an
 $SU(2)\times U(1)$ gauge theory and some
new particles or fields which spontaneously break its symmetry.
These new fields must couple to the $W$ and $Z$ bosons and thus
contribute to electroweak radiative corrections.  The constraint of the 
precision electroweak fit is that these corrections should be of the same
size as those produced by a light elementary Higgs boson.  Models in which
the Higgs boson is composite or the symmetry-breaking sector is strongly
interacting typically contain larger corrections, comparable to those of
a heavy elementary Higgs boson.  The precision electroweak 
constraint then requires
that other radiative corrections in these models cancel this contribution 
down to the small value produced by a light Higgs boson.  

About ten years
ago, at the beginning of the era of precision electroweak measurements, 
several groups studied these corrections in the simplest technicolor models of 
a strongly interacting Higgs sector.  They found that 
 the new contributions typically add to the 
heavy Higgs effect rather than cancelling it, giving an even
 stronger disagreement
with the precision data \cite{Holdom:1990tc,randall,pandt}.
To build models with a heavy Higgs boson that are 
compatible with the precision 
data, we need to find the counterexamples to this general trend.

One way to address this question is to represent the Higgs sector by the
most general possible effective Lagrangian.  Recently, a number of groups
have shown that, by adding high-dimension operators to this effective
Lagrangian, it is possible to compensate the effect of a heavy Higgs boson
and relax the upper bound on the Higgs boson 
mass \cite{BarbieriStrum,Bagger,Kolda,Chivukula}.  We believe that this
line of argument, though correct,  is incomplete.   The effective Lagrangian
description of the Higgs sector is obtained by starting with a complete
theory of electroweak symmetry breaking and integrating out the high-energy
degrees of freedom.  The full theory predicts not only the particular
operator
coefficients relevant to the Higgs mass bound but also other effects, which
might include interesting low-energy signatures~\cite{Kane:2000di}.  
Also, it could happen that
a particular set of operator coefficients cannot be produced from any 
complete theory, or may require a full theory so contrived as to be 
unacceptable.  To investigate these issues, we must go beyond the effective
Lagrangian description and ask what ingredients are needed in the full theory
to compensate the effect of a heavy Higgs boson.

In this paper, we will attempt to make general statements about the Higgs 
mass bound based on explicit models.  It is not so easy to make statements
that cover all possible models.  However, as we will review in Section 2,
the question of how to relax the constraint on the Higgs boson mass is
related to other questions about the electroweak constraints
that were raised just shortly after the inception of the precision
electroweak program ten years ago.  Considerable ingenuity has been applied
to these questions, and a substantial literature has been generated.  In this
paper, we will review this literature and extract lessons from it.

We have noticed that all explicit models proposed in the literature to 
relax the bound on the Higgs boson mass use one of three specific mechanisms,
which we will attempt to describe transparently. In Section 2, 
we will  briefly review the present
status of constraints on the Higgs boson mass from precision electroweak
interactions, using the language of $S$ and $T$ variables \cite{pandt}.
Then, in Sections 3, 4 and 5, we will discuss the three mechanisms in turn,
showing that each has a simple explanation in 
terms of the $S$, $T$ formalism. 
For definiteness, we will focus on 
models of new physics that would make a Higgs
boson of 500 GeV with Standard Model couplings consistent with the 
precision electroweak bounds.   This is a less severe criterion than that
of allowing a model with no narrow Higgs resonances and true Higgs-sector
strong interactions.

Before we begin, we have one more important introductory comment.  In models
in which electroweak symmetry breaking arises from an elementary Higgs boson,
theoretical consistency often places a stringent upper bound on the mass
of the Higgs boson which is independent of any requirement from the data.
In particular, the postulate that all interactions in Nature are weakly-coupled
up to a grand unification scale at $10^{16}$ GeV implies by itself
a very strong constraint on the Higgs boson mass~\cite{CMPP}.
  The general class of supersymmetric
grand unified theories has been 
studied exhaustively and found to give an upper bound of
205 GeV~\cite{KKW}. In addition, because of decoupling, models
which contain an 
elementary Higgs boson---even those which, like supersymmetry, contain a
huge number of new particles---typically give only small additional 
contributions to electroweak radiative corrections 
beyond the effects present
in the Standard Model.  Thus, the most familiar examples of physics 
beyond the Standard Model are compatible only with a light Higgs boson.
To allow the Higgs boson to be heavy, we must go further afield. 

%%%%%%%%%%%%%%%%%%%%%%%%%%%%%%%%%%%%%%%
\section{$S$--$T$ analysis}

The precision electroweak constraints are conveniently
represented by fitting the data to the MSM augmented by 
two `oblique' parameters, $S$ and $T$, which represent the effects of 
new physics on the $W$ and $Z$ vacuum polarization 
amplitudes \cite{pandt,other}.  We describe the method briefly.
The parameter $S$ describes weak-isospin-symmetric and $T$ describes weak
isotriplet contributions to $W$ and $Z$ loop diagrams.  Precision electroweak
observables are linear functions of $S$ and $T$.  Thus, each measurement
picks out an allowed band in the $S$--$T$ plane, and measurements of 
several processes restrict one to a bounded region in this plane.

\begin{figure}
\centerline{\epsfxsize=6.00truein \epsfbox{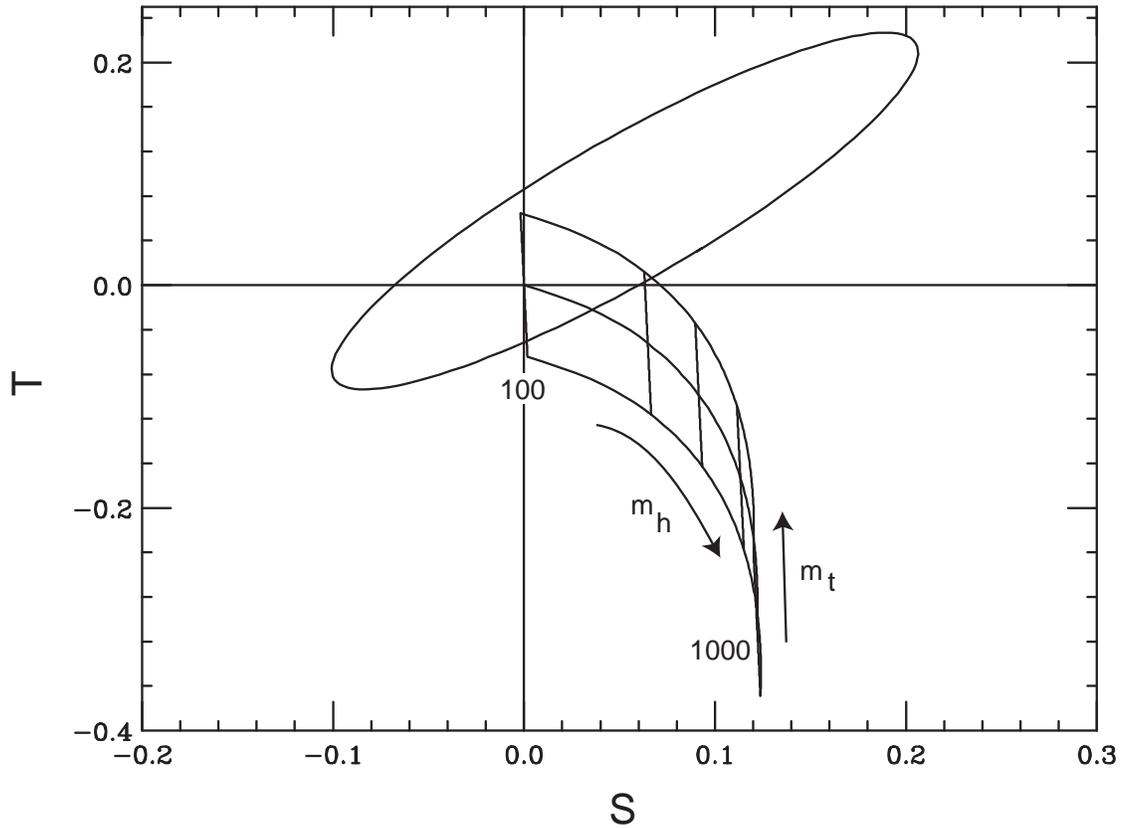}}
\vskip 0.0 cm
 \caption{Fit of the precision electroweak data to the MSM plus the 
$S$, $T$ parameters described in the text. 
 The fit is based on the values of $\mw$, $\seff$, and $\Gamma_\ell$
shown in Table~\ref{tab:data}.  The ellipse shows the 68\% two-dimensional
confidence region (1.5 $\sigma$).
  The banana-shaped figure shows the central value of a fit to the 
MSM for $m_t = 174.3 \pm 5.1$ GeV and $m_h$ varying from 100 to 
1000 GeV, with $m_h = 200$, 300, 500 GeV marked with vertical bands.
An active version of this figure can be obtained by downloading the 
additional files deposited with the eprint.}
\label{fig:STfit}
\end{figure}

By convention, the point $S = T = 0$ corresponds to the prediction of the
MSM for fixed `reference' values of the top quark and Higgs boson mass.
In this paper, we will take the reference values to be: $m_t = 174.3$ GeV
(the current central value from the Tevatron 
experiments~\cite{Partridge:1998up}) and 
$m_h = 100$ GeV. Shifts in these reference values can be compensated
by shifts in $S$ and $T$.  In Fig.~\ref{fig:STfit}, we show the 
68\% confidence contour (1.51 $\sigma$) for a current  $S$--$T$ fit.

  A fit with different reference
values of $m_t$ and $m_h$ has confidence contours of the same shape but
with a different center. The diferences from  the $S$ and $T$ values
at the original reference masses  indicate the shifts in  $S$ and $T$
that best compensate the change in the contributions from the top quark and 
Higgs boson masses.  Following 
Takeuchi, we translate the ellipse for a given $(m_t,m_h)$ back to the position
for a light Higgs boson, and then consider the translation to represent
the $(S,T)$ position associated with the new top quark and Higgs masses.
With this definition, we obtain the $(S,T)$ values shown in the figure as 
a banana-shaped grid to represent the MSM with 
$m_t = 174.3 \pm 5.1$ GeV~\cite{Partridge:1998up} 
and 
$m_h$ running from 100 GeV to 1000 GeV.  The condition $S = T = 0$ 
means `no new physics'.  From the figure, the fit strongly favors a light
Higgs boson and excludes a 500 GeV Higgs at the $5\sigma$ level.

%%%%%%%%%%%%%%%%%%%%%%%%%%%%%%%%%%%%%%%%%%%%%%%%%%%%%%%%%%%%%%%%%%%%%%
%%%%%%%%%%%%%%%%%%%%%%%%%%%%%%%%%%%%%%%%%%%%%%%%%%%%%%%%%%%%%%%%%%%%%%

The fit in Fig.~\ref{fig:STfit} uses only the values of the three 
best measured electroweak observables, $\mw$, $\seff$ (the value of
the weak mixing angle which appears in $Z^0$ decay asymmetries), and 
$\Gamma_\ell$ (the leptonic width of the $Z^0$).
Accurate analytic expressions for the Standard Model predictions for these 
quantities
have been given in \cite{gambino,marciano}.  The current values of 
these quantities are displayed in Table~\ref{tab:data}. A fit with this
data only gives
\beq
       S =  0.05  \pm  0.10   \ , \qquad   T  =  0.07  \pm 0.11  \ .
\eeq{STfitvals}
This restricted data set actually carries most of the information in a 
complete fit to the corpus of weak interaction data.  A rather sophisticated
fit presented by Swartz at the 1999 Lepton-Photon Conference \cite{morris}
gave the same errors as in \leqn{STfitvals} with a central value 
$(S,T) = (-0.04, -0.06)$. Our fit to three data points with the values 
used by Swartz gives again the same errors and a central value
$(S,T) = (0.02,-0.02)$.  The compact procedure used here makes little
difference for most of the paper, but it will considerably simplify
the analysis of Section 5.

%%%%%%%%%%%%%%%%%%%%%%%%%%%%%%%%%%%%%%%%%%%%%%%%%%%%%%%%%%%%%%%%%%%%%%%%%
\begin{table}[t]
\begin{center}
\begin{tabular}{lccr}
parameter & current value  &  $\alpha$ effect   \\ \hline
$m_W$ (GeV) & $80.434\pm0.037$ & $\pm0.003$ \\
$\seff$ & $0.23147\pm0.00017$ & $\pm 0.00006$ \\
$\Gamma_{\ell}$ (MeV) & $83.984\pm0.086$& $\pm0.028$
\end{tabular}
\end{center}
\caption{Current values of the three best measured  electroweak parameters, 
from \cite{Gurtu}.  In the fit described in the text, the 
uncertainty
in the last column \cite{marciano}, 
due to the uncertainty in $\alpha(\mz^2)$, is added
in quadrature.}
\label{tab:data}
\end{table}
%%%%%%%%%%%%%%%%%%%%%%%%%%%%%%%%%%%%%%%%%%%%%%%%%%%%%%%%%%%%%%%%%%%%%%%%%%

The earliest fits to $S$ and $T$ had central values which were substantially
negative compared to the Standard Model prediction.
It was pointed out in~\cite{pandt} that a negative value of $S$ is
especially problematic; since $S$ is the zeroth moment of a distribution
whose first momemt is positive and whose second moment is zero, $S$ will be
positive in any simple model.  This led to a number of papers on mechanisms
which generated negative $S$.  These mechanisms are directly relevant to 
our present concern.  If there is a heavy Higgs boson, it is 
clear from Fig.~\ref{fig:STfit} that we must add additional ingredients to 
the theory to compensate the effect of this Higgs boson on $S$ and $T$.
Since it is difficult to generate a negative shift in $S$, the list of 
helpful additions is severely restricted.

\section{Method A: Negative $S$}

As we have indicated in the introduction, we have exhaustively surveyed
explicit models of electroweak symmetry breaking which produce shifts in 
the $S$ and $T$ parameters.  It turns out that all such models use one of
three mechanisms to move $S$ and $T$ from the region predicted by a heavy
Higgs boson to the region preferred by the $S$--$T$ fit to data. In this
and the next two sections, we will discuss these mechanisms in turn. For
definiteness, we will consider models which contain a 500 GeV Higgs particle
and additional content associated with dynamical electroweak symmetry 
breaking.  We refer to the contributions from this additional content
as $\Delta S$, $\Delta T$.

The first method for reconciling a heavy Higgs boson with the precision
electroweak fits is to add particles whose vacuum polarization
integral shifts $S$ in the negative direction.  Typically, new 
heavy particles give a positive shift in $S$.  However, several specific
multiplets have been found which can give negative contributions to $S$.
 In this section, we will review models of this type.

Georgi \cite{georginegS} and Dugan and Randall \cite{randalldugan} 
 considered a scalar field which transforms
according to a definite representation of $SU(2) \times SU(2)$, where
the first factor is the weak interaction gauge group and the second factor
is the additional symmetry required to preserve the small value of the 
$\rho$ parameter \cite{SSVZ}. We denote the representation by $(j_L,j_R)$
according to the spin under each $SU(2)$ group.  When electroweak symmetry
is spontaneously broken, the diagonal $SU(2)$ (`custodial $SU(2)$')
is preserved, and the large
multiplet breaks up into smaller multiplets of definite spin $J$ under this
symmetry.  The smallest possible value of $J$ is $j_- = |j_L-j_R|$.  It 
turns out that, if the particles with smallest $J$ are the lightest, the 
multiplet produces negative $\Delta S$.  As long as the $SU(2)$ symmetry is
exact, the contribution to $\Delta T$ is zero.

It is interesting to ask how large a value of  $\Delta S$ can be produced
in this model.  To make a simple estimate, assume that the particle with
$J = j_-$ has the lowest mass $m$, and all other particles in the multiplet
have a common mass $M$.  Then $\Delta S$ contains a logarithm of the 
mass ratio,
\beq
    \Delta S \sim  {1\over 3\pi} X \log {M^2\over m^2} \, 
\eeq{firstnegS}
with 
\beq
    X = -  \left[ \left( {(j_+ + 1)\over (j_-+1)} \right)^2 - 1 \right]
                   {j_-(j_-+1)(2j_-+1)\over 12} \ ,
\eeq{Xval}
and $j_+ = (j_L + j_R)$.  We give a  more complete expression for  $\Delta S$
in the Appendix.  The simplest example, $(j_L, j_R) = (1,\half)$, yields
a puny coefficient
$X/3\pi = -0.024$.  Larger values can be obtained by using multiplets with
larger weak isospin.  It is important to note that the logarithm cannot be
large.  Since the mass splitting between $M$ and $m$ violates
weak isospin, this splitting 
 must be generated by electroweak symmetry breaking and 
so cannot be greater than about 100 GeV.  

\begin{figure}
\centerline{\epsfxsize=6.00truein \epsfbox{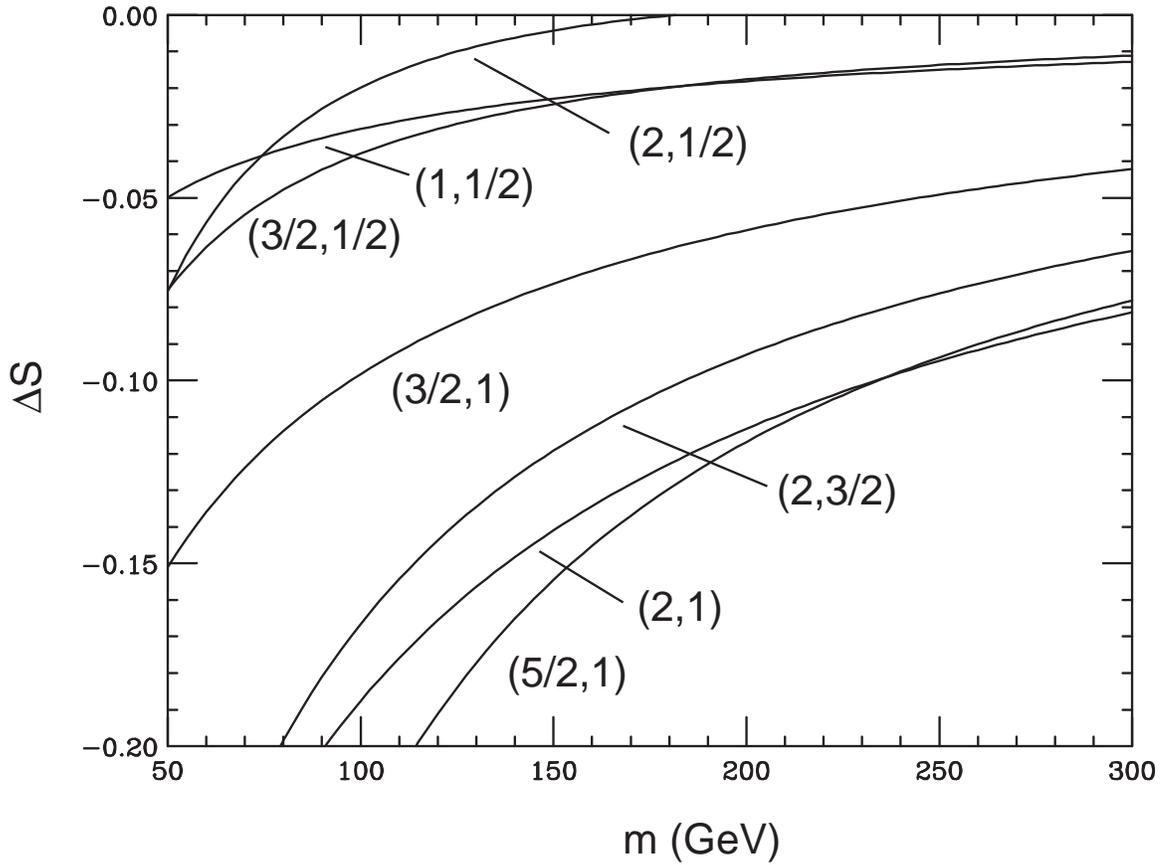}}
\vskip 0.0 cm
 \caption{Shift in $S$ induced by the vacuum polarization of various
   multiplets in the Dugan-Randall scenerio described in the text.  The
     curves assume that the lightest state in the multiplet is split in
      mass from the other states by 100 GeV.  The various Dugan-Randall
      multiplets are labeled by $(j_L,j_R)$, and the corresponding 
        shifts in $S$ are plotted against the mass of the lightest state.}
\label{fig:negS}
\end{figure}

In Fig.~\ref{fig:negS}, we plot the 
contributions to $\Delta S$ from some representative multiplets as a function
of the light mass $m$, assuming a mass splitting of 100 GeV.  The values
shown should be compared to the contribution $\Delta S = +0.11$ from a 
500 GeV Higgs boson.    Since the 
scalars involved in this mechanism couple to the weak interactions, they
will certainly be found at an $\ee$ collider that can reach their 
pair-production threshold.  From the figure, we see that it is possible
that the required
 particles might escape detection at 500 GeV $\ee$ collider,  but that 
this requires large multiplets of new particles and isospins $J > 2$.

%%%%%%%%%%%%%%%%%%%%%%%%%%%%%%%%%%%%%%%%%%%%%%%%%%%%%%%%%%%%%%%%%%%%%%
%%%%%%%%%%%%%%%%%%%%%%%%%%%%%%%%%%%%%%%%%%%%%%%%%%%%%%%%%%%%%%%%%%%%%%

We are aware of only one paper that makes use of this mechanism within a 
fully developed model of electroweak symmetry breaking.  Luty and 
Sundrum  \cite{lutysundrum} devised a set of  technicolor models in which the
pseudo-Goldstone bosons contribute a negative $\Delta S$.  However, to
obtain  $\Delta S \simeq - 0.1$ from this source,
they needed technifermions with $j_L = 2$ and pseudo-Goldstone bosons as light
as 200 GeV.  Larger values of $|\Delta S|$ could be obtained from larger
isospin multiplets.  However, these large multiplets gave compensatory
positive contributions to $\Delta S$ from the technicolor dynamics.  Other
problematical aspects of these models are also pointed out in 
\cite{lutysundrum}.

Gates and Terning \cite{gatesterning} noticed that one can obtain negative
$S$ by introducing electroweak-singlet Majorana fermions which also have
Dirac mass terms with isodoublet particles.  In this mechanism, the
contribution to $S$ again contains  a logarithm of the mass ratio of the
particles split in mass by electroweak symmetry breaking. 
The largest negative contributions to $\Delta S$ that can be obtained
by this method are of the size
\beq
        \Delta S \sim - {1\over 6\pi} \log {M_1^2\over M_2^2}
\eeq{GTform}
where $M_1$ and $M_2$ are mass eigenvalues split by electroweak symmetry
breaking.  This formula briefly led to excitement that the negative $S$
values found in electroweak fits could be explained by a chargino of 
mass 60-80 GeV \cite{ABC}.  Unfortunately, for charginos heavier than $\mz$,
the contribution decouples and vanishes as $\mz^2/M_2^2$.

Thus, to explicitly obtain negative $\Delta S$ by introducing new particles
along the lines of Dugan and Randall,
one must either introduce very large multiplets or require that
some of these particles remain light.  Light particles associated with 
this mechanism necessarily have electroweak charge and can be found at 
an $\ee$ collider.

%%%%%%%%%%%%%%%%%%%%%%%%%%%%%%%%%%%%%%%%%%%%%%%%%%%%%%%%%%%%%%%%%%%%
\section{Method B: New vectors}

The second method for reconciling a heavy Higgs boson with the precision
electroweak fits is to change the weak-interaction gauge group, adding 
heavy $Z^{0\prime}$ vector bosons.  The effects of such new bosons on 
the precision electroweak fits were studied by a number of groups in the 
early 90's \cite{LangLuo,Holdom:1991xp,Altarelli:1992fk,Altarelli:1993rk}.
Rizzo explicitly studied their effect as a method for obtaining negative
$\Delta S$ \cite{Rizzo}.  More recently, Casalbuoni \etal\
have studied the compensation of the heavy Higgs effect by new vectors 
in a model with an added gauge group 
$SU(2)\times SU(2)$  \cite{Casalbuoni}.

In all of these papers, the effects of the  $Z^{0\prime}$ is studied by 
mapping it to a shift of three variables representing the oblique electroweak
corrections, $S$, $T$, and $U$.  We find it more instructive to use a 
slightly different strategy which ignores $U$.  We will compute the shifts
in our three well-measured electroweak parameters due to the  $Z^{0\prime}$, 
fit the data to $S$ and $T$ taking these shifts into account, and see if
the resulting effect on $S$ and $T$ can compensate the effect of a heavy
Higgs boson.

For the pattern of shifts induced by a  $Z^{0\prime}$ boson, we find the 
following:  Consider a $Z^{0\prime}$ boson whose mixing with the standard
$Z^0$ is represented by the mass matrix
\beq
      m^2 =  \pmatrix{ m^2 &  \gamma \mz^2 \cr  \gamma \mz^2  & M^2} \ ,
\eeq{Zmixm}
where $\gamma$ is a parameter of order 1.  It is natural that the off-diagonal
terms are of the same order of magnitude as $\mz$ and much less than $M^2$, 
 since in typical models the heavy mass
$M^2$ results from an $SU(2)\times U(1)$-singlet expectation value,  while
both the $Z^0$ mass and the off-diagonal terms result from the expectation
values of standard Higgs fields.  The observed $Z^0$ mass is given by the lower
eigenvalue of this matrix, $\mz^2 = m^2 ( 1 -  \delta )$,
and the physical $Z^0$ contains an admixture $\xi$ of the original  
$Z^{0\prime}$, where
\beq
       \delta = \gamma^2 {\mz^2\over M^2} \ , \qquad  
              \xi =  \gamma {\mz^2\over M^2} \  ,
\eeq{dgdefs}
to leading order in $(\mz^2/M^2)$.  Let the current coupling the $Z^{0\prime}$
to leptons $\ell^-$ have the form
\beq
    \Delta \L =  g' Z^{0\prime}_\mu \left\{ \bar \ell_L \gamma^\mu q_L \ell_L
                   + \bar \ell_R \gamma^\mu q_R \ell_R \right\} \ .
\eeq{Zcurrents}
Then the  $Z^{0\prime}$ induces the shifts
\beqa
     \Delta \mw & = & 57.\, \delta \quad \mbox{(GeV)}  \CR
    \Delta \seff & = & -0.33\, \delta + 0.22\, q_L\xi + 0.26\, q_R\xi\CR
    \Delta \Gamma_\ell & = & 100\, \delta -  170\, q_L\xi + 150\, q_R\xi
                                       \quad \mbox{(MeV)} \ .
\eeqa{Zpshifts}
Symbolic versions of these expressions are given in the Appendix.

\begin{figure}[p]
\centerline{\epsfxsize=6.00truein \epsfbox{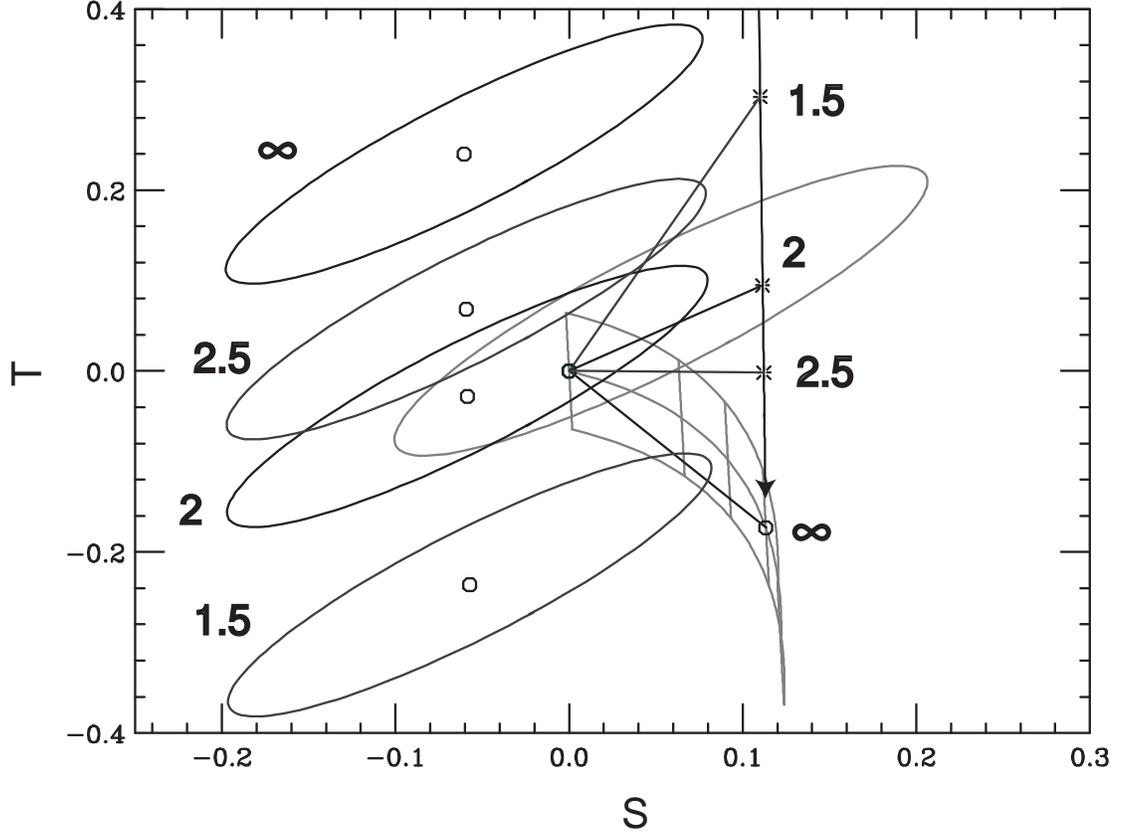}}
\vskip 0.0 cm
 \caption{Fit of the precision electroweak data to the MSM with 
 $m_h = 500$ GeV and shifts of the electroweak parameters due to a 
 $Z^{0\prime}$, plus the effects of the
$S$, $T$ parameters. The four darker ellipses correspond to fits with 
$M = 1.5$, 2.0, 2.5 TeV and $\infty$.  The lighter ellipse and the grid
are those plotted in Fig.~\ref{fig:STfit}. This diagram shows how the
centers of the 
various fits with different values of $M$  (symbolized by  $\circ$)
can be plotted as shifts of $(S,T)$
with respect the Standard Model ellipse (symbolized by $*$).  These shifts
represent the combined contribution of the $Z^{0\prime}$
and the heavy Higgs boson, and fall on a line which tends to the 
heavy Higgs boson prediction for $M \to \infty$. 
 We see almost complete compensation of the 
heavy Higgs boson effect for $M \sim 2$ TeV.}
\label{fig:STfitZpC}
\end{figure}

To demonstrate the effect of these shifts, consider first the simple
case $q_L = q_R = 0$, and take $\gamma = 1$. (This last choice is conservative,
since typically $\gamma$ is of order $\sstw$.)   Now set the Higgs boson mass
to 500 GeV, add to the MSM prediction the shifts shown in \leqn{Zpshifts}, 
and fit for $S$ and $T$.  The result is the set of contours shown in 
Fig.~\ref{fig:STfitZpC}.  We see almost complete compensation of the 
heavy Higgs boson effect for $M \sim 2$ TeV.  The figure shows how we 
would plot this compensation as a translation of the center of the fit,
in the same way that we plotted the $(S,T)$ contributions from shifts in 
$m_t$ and $m_h$ within the MSM.

In principle, the fit might have become worse as the
$Z^{0\prime}$ pulls the three variables in directions that are not possible
within the MSM.  To show that this does not happen significantly, we 
have plotted the various new ellipses at the same $\chi^2$ value as the 
reference ellipse copied from Fig.~\ref{fig:STfit}.  If a third variable $U$
were required, the contours would become smaller for low $Z^{0\prime}$ masses,
but this clearly does not happen.  In this special case with only
$Z-Z'$ mixing, the 
effect of the $Z^{0\prime}$ is actually completely described by a shift of
$T$, $\Delta T = \delta/\alpha$.  However, it is true in the other examples
we have studied that the main effect of the  $Z^{0\prime}$ is to shift
the center of the $(S,T)$ fit while maintaining a fit with reasonable 
$\chi^2$.

%%%%%%%%%%%%%%%%%%%%%%%%%%%%%%%%%%%%%%%%%%%%%%%%%%%%%%%%%%%%%%%%%%%%%%
%%%%%%%%%%%%%%%%%%%%%%%%%%%%%%%%%%%%%%%%%%%%%%%%%%%%%%%%%%%%%%%%%%%%%%

\begin{figure}
\centerline{\epsfxsize=6.00truein \epsfbox{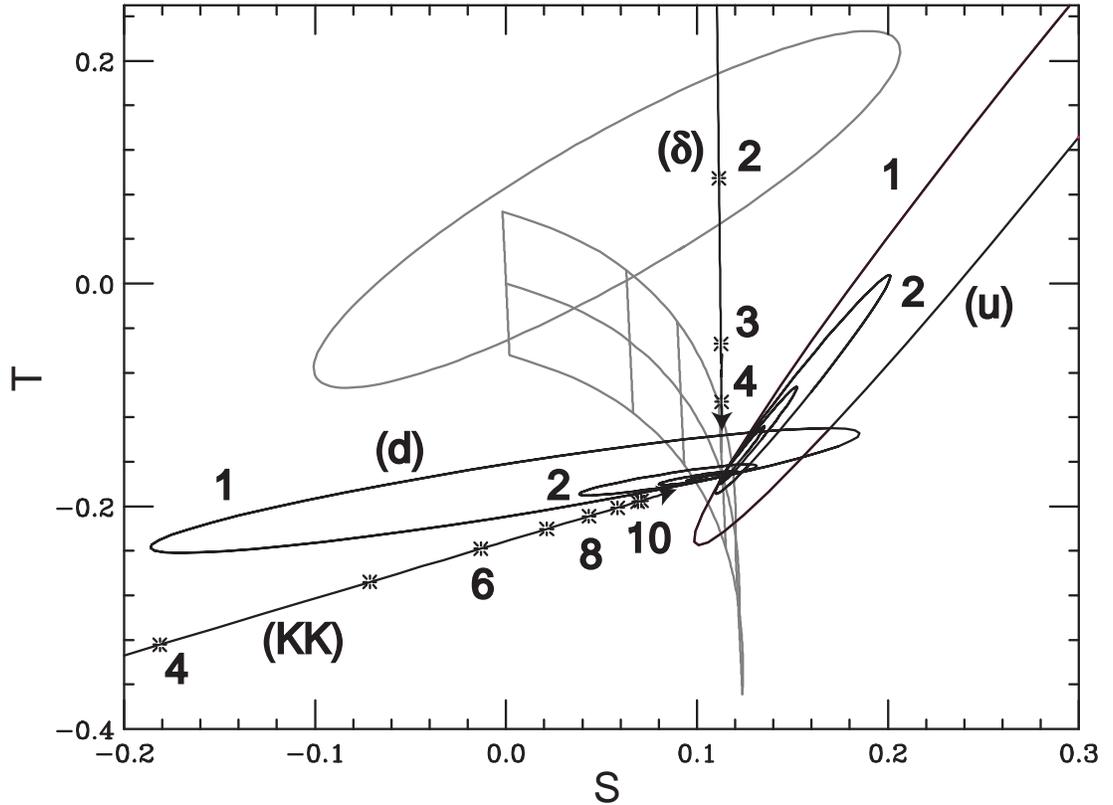}}
\vskip 0.0 cm
 \caption{Contributions to $S$ and $T$ from a Higgs boson with 
 $m_h = 500$ GeV, plus a heavy 
 $Z^{0\prime}$.  The contributions are computed and displayed as 
indicated in Fig.~\ref{fig:STfitZpC}.  Four different models are considered:
($\delta$):  model of  Fig.~\ref{fig:STfitZpC}, with $\gamma =1$, 
$q_{L,R} = 0$;
(u): rank-1 $E_6$ models with mixing due to a Higgs field $H_u$; (d):
 rank-1 $E_6$ models with mixing due to a Higgs field $H_d$; (KK):  
extra-dimension model of ref. \cite{rizzowells}.  The numbers indicate 
the values of the  $Z^{0\prime}$ mass $M$, always in TeV, and the
star symbols represent the $(S,T)$ shifts, as in Fig.~3, for the variously
labeled values of $M$.  For the $E_6$ models, there are two parameters
to vary, the mass $M$ and the mixing angle $\theta$.  For these 
models, we have plotted the contours swept out as one changes the 
mixing angle for fixed values of  $M$.
All of the $Z^{0\prime}$ predictions tend to the 500 GeV MSM 
point as $M \to \infty$.}
\label{fig:STwZ}
\end{figure}

Using this formalism, we can investigate the region of parameters for any
 $Z^{0\prime}$ model in which the shifts due to the $Z^{0\prime}$ compensate
those of a heavy Higgs boson.  In Fig.~\ref{fig:STwZ}, we show the
results for the fit 
centers as a function of the parameters of the  $Z^{0\prime}$,
for the model described in 
 the previous paragraph and for several other models
from the literature.  A commonly discussed
class of  $Z^{0\prime}$ models are the rank-1 $E_6$ models, obtained
by considering the  $Z^{0\prime}$ to be an arbitrary linear combination of 
the two $U(1)$ bosons in $E_6$ that are orthogonal to the bosons of 
$SU(3)\times SU(2)\times U(1)$.  The $SO(10)$  $Z^{0\prime}$ and the
`superstring-inspired'  $Z^{0\prime}$ are particular cases of these models.
The predictions of these models for precision electroweak parameters 
depend through $\xi$ on the quantum numbers of the Higgs field responsible
for $Z$--$Z'$ mixing.  This Higgs field could be either an $H_u$, with
$I$ = $\half$, $Y$ = $\half$, or an $H_d$, with
$I$ = $\half$, $Y$ = $-\half$, or the mixing could receive contributions
from Higgs fields of both types.  Explicit formulae for the various
contributions from  these  models 
are given in the Appendix.
 In Fig.~\ref{fig:STwZ}, we plot the 
compensation as a function of the $E_6$ mixing angle for each of the two 
extreme cases, for fixed values of $M$.  In these models,  partial 
compensation is possible only for relatively low values of $M$,  below
1.5 TeV. 

Figure 4 contains a substantial amount of detailed information, but it also
contains two simple messages.  First, it is possible within the space of 
$Z'$ models to arrange shifts $(\Delta S, \Delta T)$ that move the precision
electroweak fit in almost any direction.  Second, in any given model, 
these shifts can be large
enough to influence the conclusion about a heavy Higgs boson only if the 
mass of the $Z'$ is small enough.

%%%%%%%%%%%%%%%%%%%%%%%%%%%%%%%%%%%%%%%%%%%%%%%%%%%%%%%%%%%%%%%%%%%%%%
%%%%%%%%%%%%%%%%%%%%%%%%%%%%%%%%%%%%%%%%%%%%%%%%%%%%%%%%%%%%%%%%%%%%%%

The final model shown in Fig.~\ref{fig:STwZ} is a model involving extra 
space dimensions \cite{rizzowells}.  This is an example of a number of 
models presented recently in which gauge fields live in a higher-dimensional
space that is
compactified to our ordinary $3+1$ 
dimensions \cite{Antoniadis,Pomarol,Delgado,Strumia}.  
In these models, the new vectors
are the Kaluza-Klein excitations of the MSM gauge bosons.  

These models with extra dimensions are interesting to this study because
one of the simplest
of these theories---with gauge fields in 5-dimensions and 
the Higgs boson and fermions in 4-dimensions---exhibits compensation 
between the effects of the new vector bosons and the heavy Higgs boson.
If we denote the mass of the first KK excited state of the
gauge bosons by $M_{KK}$, the electroweak observables are altered in 
this model by
\beqa
\Delta m_W & = & 87.0 \frac{m_Z^2}{M_{KK}^2} \quad \mbox{(GeV)} \CR
\Delta \seff  & = & -1.09 \frac{m_Z^2}{M_{KK}^2} \CR
\Delta\Gamma_{\ell} & = & -220 \frac{m_Z^2}{M_{KK}^2} \quad \mbox{(MeV)} \ .
\eeqa{KKshifts}
The effect of these changes is to move the $(S,T)$ value to a region
of the plane where it is less constrained. Though the 68\% C.L. ellipse
does not intersect the KK line in Fig.~4, the expanded ellipse corresponding
to the 99\% C.L. reaches into the lower left-hand corner of the plot and 
intersects the KK line for $M_{KK} \sim 3$ TeV.
A detailed global fit done in the summer of 1998~\cite{rizzowells}
indicated that the 95\% C.L. upper bound on the Higgs boson mass
could reach as high
as 300--500 GeV for $M_{KK}$ in the range 3--5 TeV. With the latest 
experimental numbers, we find that the upper limit on the Higgs boson mass
is still relaxed in this model, though it does not extend beyond 300 GeV.
 
Another interesting property of this model is that the coupling of the
new sector is large. By this we mean both that a large number of new
states participate and that the couplings of individual states are 
larger by a factor $\sqrt{2}$ than those of the corresponding MSM bosons.
These features allow compensation for a mass of the lightest new vector 
about twice as high as the mass of the single $Z^{0\prime}$ boson in 
the $E_6$ models considered above.

A variation on this scheme is suggested in \cite{kribs}, which considers
a higher-dimen\-sion\-al model with a low quantum gravity scale.  In this case,
the precision electroweak corrections are distorted by the effect of the
radion, a scalar degree of freedom from the gravity sector. The corrections
involve the value of the underlying Planck scale $M$ and the Higgs field
coupling to curvature through a term $\half \xi R h^2$.  The model allows
compensation of the electroweak corrections and a heavy Higgs boson,
but only when $M$ or $M/\xi$ is close to 1 TeV.

The common feature of these four models, and of other $Z^{0\prime}$ models
we have studied, is that the values of the  $Z^{0\prime}$ mass needed to 
compensate the effect of a heavy Higgs boson is well within the reach of 
next-generation colliders.  The LHC should be able to find a  $Z^{0\prime}$ 
as a narrow resonance for masses up to 4 TeV.  A 500 GeV $\ee$ linear collider
can see the effect of the  $Z^{0\prime}$ as a perturbation of the cross 
section for $\ee\to f\bar f$, with similar  sensitivity in the 
$Z^{0\prime}$ mass \cite{Rizzo:1998cc}.  The information from proton and
electron colliders is complementary, and a complete picture of the 
$Z^{0\prime}$ is obtained by combining the two sets of measurements.
In the case of the extra-dimension model, the mass of the first new vector
excitation is predicted to be higher.  However, in this model, 
the larger couplings give 
enhanced sensitivity, up to 6 TeV for the LHC and above 10 TeV for a 500 GeV
$\ee$ linear collider, so the general conclusion applies to this model as well.
 
\section{Method C: Positive $T$}

In both methods of compensating a heavy Higgs boson that we have 
discussed so far, the compensation leads to new physics signatures that
should be observed at next-generation $pp$ and $\ee$ colliders.  However, 
there is one further compensation strategy that can evade this requirement.
Looking again at Fig.~\ref{fig:STfit}, we see that it is possible to 
bring a model with a heavy Higgs boson back into reasonable agreement
with the precision electroweak fit without changing $S$ at all, by 
adding new particles that lead to positive $\Delta T$.  For example, 
the shift $\Delta S = 0$, $\Delta T = 0.3$ due to new physics 
brings a model with a 500 GeV
Higgs boson within 1 sigma of the central value.

Most models with new physics produce
a nonzero, positive $\Delta T$ \cite{EinhornJones}. In fact, the contribution
to $\Delta T$ can easily be of order 1.  Particles with mass much larger than
1 TeV can contribute to $\Delta T$ if their masses have an up-down
 flavor asymmetry.  The contribution is of the order of 
\beq
          \Delta T \sim   {m_U^2 - m_D^2 \over m_U^2 + m_D^2}  \ .
\eeq{Tcontrib}
Even though $m_U - m_D$ can be at most of order 100 GeV because it must arise
from electroweak symmetry breaking, this contribution can easily be large
enough to compensate the effect of a heavy Higgs boson for values of the $U$
and $D$ masses that are inaccessible to any collider.

Several recently proposed models allow a heavy Higgs boson to make use of
this mechanism.  The first is the `topcolor seesaw' of Dobrescu and 
Hill \cite{topcolorseesaw}.
In this model, the new physics needed to break electroweak symmetry arises
from a heavy, weak-$SU(2)$-singlet fermion $\chi$. In the
simplest topcolor seesaw model, one finds \cite{georgistccsaw}
\beq
  \alpha \Delta T =  {3\over 16\pi^2} 
{g_{tc}^4\over 4} {v^2 \over m_\chi^2}
\left[ 1 +  2 {\lambda_t^2\over g_{tc}^2} \log{m_\chi^2\over m_t^2} \right]\ ,
\eeq{Tintccsaw}
where $g_{tc} \sim 3$ is the topcolor coupling, $\lambda_t = 1$ is the
top quark Yukawa coupling, $v = 246$ GeV is the weak interaction scale,
and $m_\chi$ is the mass of a new heavy fermion.  For $m_\chi =  1$ TeV,
this expression gives $\Delta T = 7.2$.  However, it is permissible in this
model to raise $m_\chi$ arbitrarily, although very high $m_\chi$ requires
fine-tuning of the underlying parameters.
   For $m_\chi = 5$ TeV, we find $\Delta T = 0.3$, which gives a
reasonable fit to 
the precision electroweak data with heavy Higgs boson. It is argued in 
\cite{topcolorseesaw} that this choice does not yet require fine-tuning. 
A mapping of the 
ellipse in Fig.~\ref{fig:STfit} into the $m_\chi,m_h$ plane gives the 
interesting contour seen in Fig. 7 of \cite{georgistccsaw}.  Similar
behavior is seen in the `topflavor' model of \cite{Tait}.  

A recent paper by  Chankowski \etal\ \cite{Chankowski:2000an}
argues that the
two-Higgs-doublet model can be made consistent with the electroweak fits
for a Higgs boson mass of 500 GeV.  The strategy of this paper is to adjust
the Higgs spectrum to give the required positive contribution to the 
$\rho$ parameter ($\Delta\rho = \alpha \Delta T$); the model gives only a 
tiny shift in $S$. A recent paper by He \etal\ \cite{He} suggests adding
a fourth generation of quarks and leptons.  The additional fermions increase
$S$, but even $S\sim 0.2$ may be accomodated by choosing the mass spectrum
to give an appropriate value of $T$.

Technicolor models can also be  made consistent with the 
precision electroweak fits through this strategy.  Most technicolor models 
lead to values of $S$ larger than the value for a 1000 GeV Higgs 
boson,  $S > 0.12$ \cite{pandt}.  In typical cases, 
the values of $S$ and $T$ are positive and of order 1.   Models have been
proposed in which the technicolor enhancements  to 
$S$ and $T$ are of order 0.1 or smaller \cite{sundrumetc,apptern,postmodern}.
But in all models that have been studied, except for \cite{lutysundrum} cited
above, the lower bound for $S$ still applies.
Still, it is possible to construct a technicolor model that is 
consistent with the electroweak data in spite of this bound, by including
enough weak isospin breaking to give a small positive correction to $T$.
Such a model would, for example, have  $S \sim 0.15$, $T \sim 0.2$.
Models of this type would not have a visible Higgs boson and might not
contain any new particles below the first techi-rho 
resonance at about 2~TeV~\cite{Barklow}.

It is important to note that the models we have discussed in the last few
paragraphs are minimal ones that represent the worst-case scenarios for 
the colliders of the next generation.  More typical and realistic
 models of  topcolor and technicolor contain additional ingredients that
form  the basis for further experimental signatures.  These include
additional gauge groups~\cite{hillgeorgi} or extra space 
dimensions~\cite{bogdancheng} in the case of topcolor and light
techni-pions and techi-rho states~\cite{newlane} in the case of technicolor.

\begin{figure}
\centerline{\epsfxsize=6.00truein \epsfbox{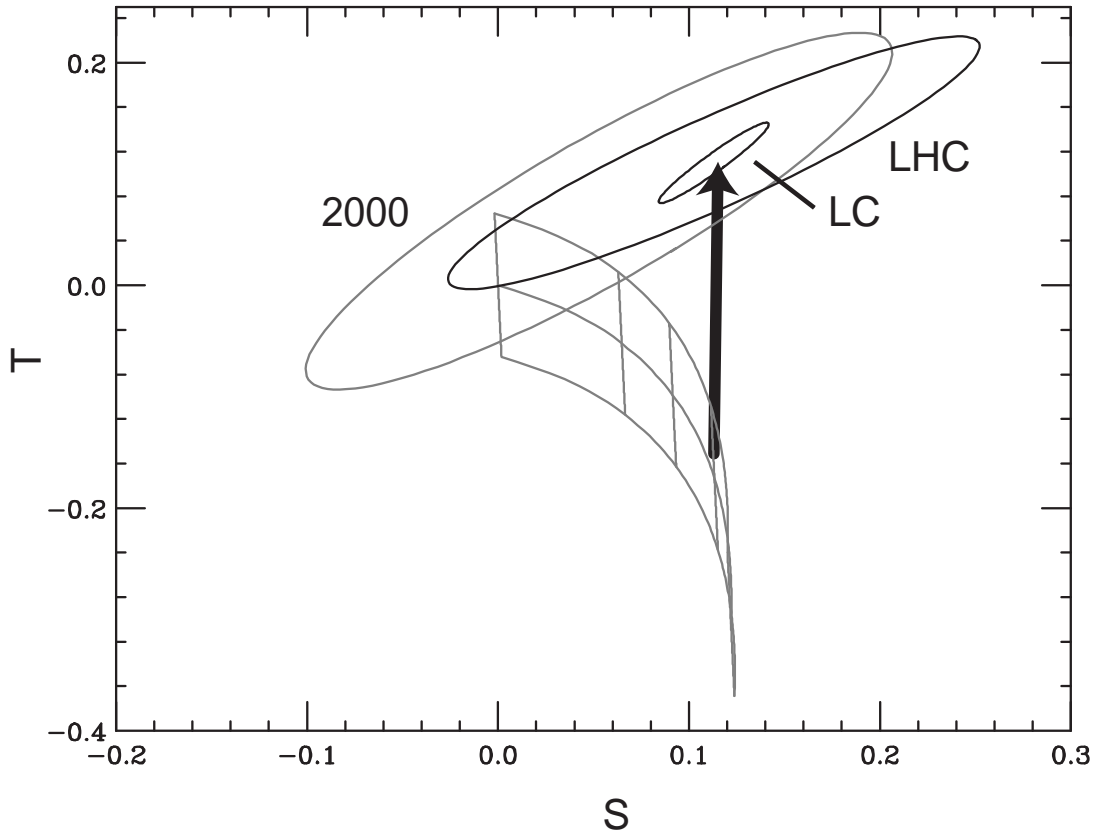}}
\vskip 0.0 cm
 \caption{Future improvements in the determination of precision electroweak
parameters.  The lighter ellipse and grid
are those plotted in Fig.~\ref{fig:STfit}. The heavier ellipses, 
both centered at $(S,T) = (0.11, 0.11)$,  correspond to an improved
$W$ mass measurement
with an error of 15 MeV, as would be expected from the LHC, and measurements
of  $\mw$, $\seff$, and $\Gamma_\ell$ with errors of 6 MeV, 0.00002, and 
0.04 MeV, respectively, as would be expected from the precision electroweak
program at an $\ee$ linear collider \cite{TESLAEW,Wilson,Erler,Rowson}.}
\label{fig:better}
\end{figure}

However, even those models using this strategy which predict little 
or no new physics at
the next generation of colliders will be clearly distinguishable from the 
MSM by improved precision electroweak measurements.  Foreseeable improvements
in the precision electroweak fit are shown in Fig.~\ref{fig:better}.
The larger contour shows the effect of a measurement of $\mw$ to 15 MeV, as
might be expected from the LHC~\cite{Gianotti:2000tz}.  
  The smaller contour shows the effect of 
the precision measurements expected from a high-luminosity $\ee$
 linear collider
run at the $Z^0$ and at the $W^+W^-$ threshold: $\mw$ to 6 MeV, $\seff$ to 
0.00002, $\Gamma_{\ell}$ to 0.04 MeV \cite{TESLAEW,Wilson,Erler,Rowson}.  
With this latter set of measurements, the point in the $(S,T)$ space
favored by this strategy is separated from prediction of 
the MSM with a light Higgs boson by more than 5 sigma.  Thus, these 
measurements would clearly prove the presence of new physics and would 
indicate the route by which today's precision electroweak constraint is 
evaded.

%%%%%%%%%%%%%%%%%%%%%%%%%%%%%%%%%%%%%%%%%%%%%%%%%%%%%%%%%%%%%%%%%%%%%%
%%%%%%%%%%%%%%%%%%%%%%%%%%%%%%%%%%%%%%%%%%%%%%%%%%%%%%%%%%%%%%%%%%%%%%

\section{Conclusions}

The precision electroweak data are consistent with the Minimal
Standard Model only if the Higgs boson mass is very low,
$m_h< 170$ GeV at 95\% C.L.  This result would predict that the colliders of 
the next generation, possibly including the upgraded Tevatron,
 will be able to discover
and study the Higgs boson.  However, if the Minimal 
Standard Model is not correct,
there are scenarios in which  new physics contributions conspire
with a heavier Higgs boson to allow agreement with the precision electroweak
data.  

In this paper, we have argued that, despite this, one cannot freely assume 
that the Higgs boson is 
heavy in the face of the precision electroweak constraint.  The particular
new physics that compensates the effect of a heavy Higgs boson  has a price,
and, to evade the MSM constraint on the Higgs boson mass, one must be 
prepared to pay it.

Many popular models of physics beyond the Standard Model do not allow a 
heavy Higgs boson at any price.  Supersymmetric grand unified theories are
an example.  Among models that allow a heavy Higgs boson in principle,
any successful model must
introduce new physics that can perturb the precision electroweak observables
in the correct direction by a sufficiently large amount.  In this paper, 
we have 
made that statement precise, using the $S$, $T$ formalism, and we have 
reviewed the various strategies suggested in the literature.  In fact, the
entire literature to date is exhausted by only three strategies, which we
have described in detail.

Two of these strategies, method A, which gives new contributions of 
negative $\Delta S$, and method B, which introduces new vector bosons, 
have distinctive signatures that should be observed at the next $\ee$
linear collider.  Method A requires new light bosons or fermions with 
electroweak charge.  Method B requires that the new vector bosons be 
sufficiently light to create large perturbations of the cross sections for
$\ee$ annihilation to fermion pairs.  In fact, models using this strategy
create a very interesting physics scenario, in which  the new vector
particles are also directly observable as resonances at the LHC.

The third strategy, method C, is less dramatic, and specific models exist
in which no new particles beyond the MSM are observable either at an $\ee$
linear collider or at the LHC.  However, this strategy leads to a 
prediction for the precision electroweak parameters that is distinctive, 
and that can be distinguished from the MSM prediction with a high level of
confidence by the improved level of precision in the electroweak parameters
that a next-generation $\ee$ linear collider should achieve.

We cannot close off the idea that the Higgs boson is very heavy on purely 
theoretical grounds.  But we have 
emphasized in this paper that models in which the Higgs boson is heavy 
have specific properties which must be taken into account in any discussion
of future experimental prospects.  In particular, these models generally 
lead in their own way to an interesting experimental program for the 
next-generation colliders. 
% A 500 GeV $\ee$ linear collider has an 
%important---and, in some cases, crucial---role in untangling the new physics
%that these models present.

\Acknowledgements
We are grateful to Gordy~Kane and Ken~Lane for 
inspiration, to
Bogdan~Dobrescu, Andreas Kronfeld,
 and the participants of the Berkeley 2000 Linear Collider
Workshop for valuable discussions, and to Jonathan~Bagger and Morris~Swartz
for useful correspondence.  

\appendix

\section{Appendix}

In this appendix, we present various explicit formulae that are used 
in the text.

In Section 3, we analyzed the Dugan-Randall models \cite{randalldugan}
in the most favorable case in which the scalar particle with smallest 
$J$ has a low mass $m$ while the other particles in the multiplet have a 
large mass $M$.  In \leqn{firstnegS},
we quoted only the leading logarithm in the formula for 
$\Delta S$.  The complete formula for this case is
\beq
   \Delta S =  
         {1\over 3\pi} \left[ X \log {M^2\over m^2} + 2 X' B(m,M)\right]\ , 
\eeq{DRDS}
where 
\beqa
      X &=&  \left[1-  \left({(j_+ + 1)\over (j_-+1)} \right)^2\right]
                   {j_-(j_-+1)(2j_-+1)\over 12} \CR
      X' &=& \left[1 +  \left( {(j_+ + 1)\over (j_-+1)} \right)^2
                          - {j_+(j_+ +2) + j_-^2\over j_- (j_-+1)} \right]
                   {j_-(j_-+1)(2j_-+1)\over 12} \CR
      B(m,M) &=& - {m^4(m^2 - 3 M^2) \over (M^2 -m^2)^3}\log {M^2\over m^2}
                    + {5 M^4 - 22 M^2 m^2 + 5 m^4\over 6 ( M^2 - m^2)^2} \ .
\eeqa{DRparts}

In Section 4, we analyzed the effect of a $Z^{0\prime}$ boson on the 
best-measured precision electroweak observables.  In \leqn{Zpshifts},
we quoted numerical formulae for the shifts in $\mw$, $\seff$, and 
$\Gamma_\ell$ induced by a  $Z^{0\prime}$ in terms of the parameters
$\delta$, $\xi$, $q_{L,R}$ defined in \leqn{dgdefs}, \leqn{Zcurrents}.
The corresponding analytic formulae are
\beqa
     \Delta \mw & = &\half {c^2\over c^2 - s^2} \mw \delta \CR
    \Delta \seff & = &  - {s^2c^2\over c^2 -s^2} \delta + 
                       s \xi (q_R (1-2s^2) + 2 s^2 q_L) \CR
    \Delta \Gamma_\ell & = & \Gamma_\ell \bigg\{ \left({1-2s^2\over s^2 c^2} + 
 {4 (1-4s^2)\over 1 - 4 s^2 + 8s^4}\right) {s^2c^2\over c^2 -s^2}\delta\CR
  & & \hskip 0.5in
- {4\over  1 - 4 s^2 + 8s^4} s \xi (q_L (1-2s^2) - 2 s^2 q_R) \bigg\} \ ,
\eeqa{Zpshiftsa}
where $s = \sin\theta_w$, $c = \cos\theta_w$.
In the rank-1 $E_6$ models \cite{LangLuo},
\beqa
   q_L &=&  \cos\theta {3\over 2\sqrt{6}} 
                   + \sin \theta {1\over 6}\sqrt{{5\over 2}}   \CR 
   q_R &=&  \cos\theta {1\over 2\sqrt{6}} 
                   - \sin \theta {1\over 6}\sqrt{{5\over 2}} \ ,
\eeqa{Zpparams}
where $\theta$ is the mixing angle between the two $U(1)$ bosons, defined
so that $\theta = 0$ corresponds to the $SO(10)$ boson $\chi$ and 
 $\theta = \pi/2$  to the $E_6$ boson $\psi$.  The expressions
for $\delta$ and $\xi$ require a parameter $\gamma$, which depends on the 
quantum numbers of the Higgs boson responsible for $SU(2) \times U(1)$
breaking and 
$Z^0$--$Z^{0\prime}$ mixing.
In general, we would expect  both Higgs fields $H_u$ and $H_d$ to obtain
vacuum expectation values, which are conventionally written
\beq
    \VEV{H_u^0} = {v\over \sqrt{2}} \sin\beta \qquad
    \VEV{H_d^0} = {v\over \sqrt{2}} \cos\beta \ ,
\eeq{HUDvals}
with $v = 246$ GeV. Then 
\beq
    \gamma = 2 s \sin^2\beta
    (\cos\theta {1\over \sqrt{6}} - \sin\theta\sqrt{5\over 18}) + 
 2 s \cos^2 \beta (\cos\theta {1\over \sqrt{6}} +\sin\theta\sqrt{5\over 18})\ .
\eeq{firstgamma}
The cases (d) and (u) shown in Fig.~\ref{fig:STwZ} correspond to the cases
$\beta = 0$ and $\beta = \pi/2$, respectively.

\end{document}